\def\rn{\noindent\parshape 2 0truecm 8.5truecm 0.3truecm 8.2truecm}
\def\rn{}
\def\nn#1 #2{#2. #1}				
\def\nnn#1 #2 #3{#2. #3. #1}			
\def\nnnn#1 #2 #3 #4{#2. #3. #4 #1}		
\def\nnnnn#1 #2 #3 #4 #5{#2. #3. #4 #5. #1}	
\def\dualand{ and\hbox{ }}				
\def\multiand{, and\hbox{ }}				
\def\rf#1;#2;#3;#4;#5 {{\frenchspacing\par\rn#1, #3 {\bf #4}, #5 (#2). \par}}
\def\rg#1;#2;#3;#4;#5;#6 {{\frenchspacing\par\rn#1, #3 {\bf #4}, #5 (#2). \par}}
\def\rfbook#1;#2;#3;#4;#5 {{\frenchspacing\par\rn#1, {\it #3} (#5, #4, #2).\par}}
\def\rfprep#1;#2;#3 {{\par\frenchspacing\rn#1, #3 (#2).\par}}

\def\muK{{\rm \mu K}}

\def\expec#1{\langle#1\rangle}

\def\etal{{\frenchspacing\it et al.}}
\def\ie{{\frenchspacing\it i.e.}}
\def\eg{{\frenchspacing\it e.g.}}
\def\etc{{\frenchspacing\it etc.}}

\def\beq#1{\begin{equation}\label{#1}}
\def\eeq{\end{equation}}
\def\beqa#1{\begin{eqnarray}\label{#1}}
\def\eeqa{\end{eqnarray}}
\def\eq#1{equation~(\ref{#1})}

\def\fig#1{Figure~\ref{#1}}
\def\Fig#1{Figure~\ref{#1}}

\def\sec#1{Section~\ref{#1}}

\def\spose#1{\hbox to 0pt{#1\hss}}
\def\simlt{\mathrel{\spose{\lower 3pt\hbox{$\mathchar"218$}}
     \raise 2.0pt\hbox{$\mathchar"13C$}}}
\def\simgt{\mathrel{\spose{\lower 3pt\hbox{$\mathchar"218$}}
     \raise 2.0pt\hbox{$\mathchar"13E$}}}
\def\simpropto{\mathrel{\spose{\lower 3pt\hbox{$\mathchar"218$}}
     \raise 2.0pt\hbox{$\propto$}}}

\def\ed{\end{document}}

\def\draft{
}

\def\arcdeg{^\circ}

\def\l{\ell}

\def\r{{\bf r}}
\def\x{{\bf x}}

\def\z{{\bf z}}

\def\xt{\tilde{\x}}

\def\I{{\bf I}}
\def\L{{\bf L}}

\def\N{{\bf N}}

\def\SS{{\bf S}}

\def\Sig{{\bf\Sigma}}

\def\tr{\hbox{tr}\,}

\def\xwhite{\x_{\rm white}}
\def\xblue{\x_{\rm blue}}

\def\Sblue{{\bf\Sigma}_{\rm blue}}
\def\L{\bf L}

\documentstyle[prd,aps,epsf]{revtex}
\draft
\begin{document}
\twocolumn[\hsize\textwidth\columnwidth\hsize\csname@twocolumnfalse\endcsname



\title{The CMB power spectrum at $\l=30-200$ from QMASK}

\author{Yongzhong Xu, Max Tegmark, Angelica de Oliveira-Costa}

\address{Dept. of Physics, Univ. of Pennsylvania, Philadelphia, PA 19104;
  xuyz@physics.upenn.edu}

\date{Submitted to Phys. Rev. D May 1 2001, accepted Jan 15 2002}

\maketitle 

\begin{abstract} 
We measure the cosmic microwave background (CMB) power spectrum
on angular scales $\l\sim 30-200$ ($1^\circ-6^\circ$)
from the QMASK map, 
which combines the data from the
QMAP and Saskatoon experiments. Since the accuracy of recent 
measurements leftward of the first acoustic peak is
limited by sample-variance, the large area of the QMASK map 
(648 square degrees) allows us to place among the sharpest constraints
to date in this range, in good agreement with
BOOMERanG and (on the largest scales) COBE/DMR.
By band-pass-filtering the QMAP and Saskatoon maps, we are able to 
spatially compare them scale-by-scale to check for beam- and pointing-related
systematic errors. 
\end{abstract}

\bigskip

] 


\section{INTRODUCTION} 
\label{IntroSec}

After the discovery of large-scale 
Cosmic Microwave Background (CMB) fluctuations 
by the COBE satellite \cite{Smoot92}, experimental groups have
forged ahead to probe ever smaller scales.
Now that TOCO \cite{Torbet99,Miller99}, 
BOOMERanG \cite{deBernardis00}
and 
Maxima \cite{Hanany00} 
have convincingly
measured the location and height of the first acoustic peak,
attention is shifting to still smaller scales to resolve outstanding 
theoretical questions. 
For instance,  the structure of the second and third peaks 
constrains the cosmic matter budget\footnote{
After this paper was first submitted, accurate new measurements
of these higher peaks were reported by the 
Boomerang \cite{Netterfield01}, DASI \cite{Halverson01} and 
Maxima \cite{Lee01} teams, so we have added comparisons with
these results below.
}.
However, it remains important to improve measurements on
larger angular scales as well, both for measuring cosmological 
parameters and to cross-validate different experiments against potential 
systematic errors. This is the goal of the present paper.

\begin{figure}[tb]
\centerline{\epsfxsize=8.5cm\epsffile{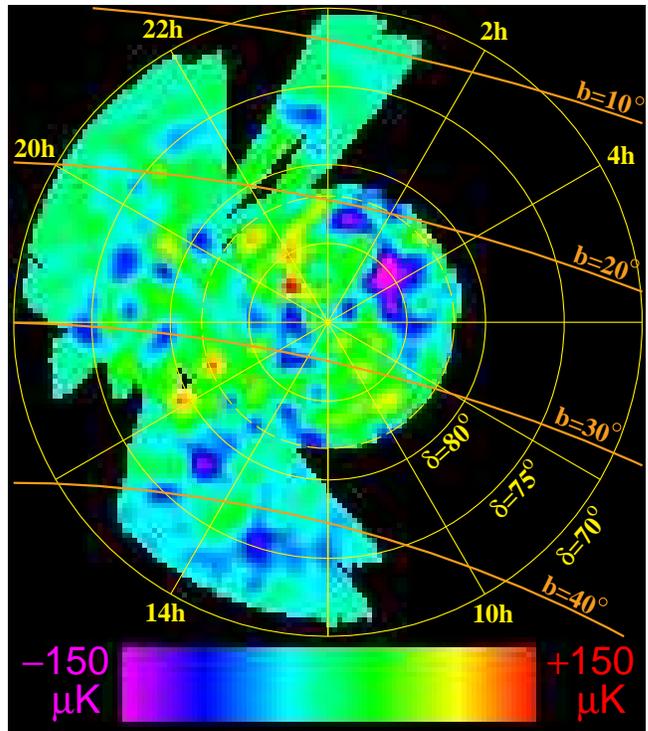}}
\smallskip
\caption{\label{newqmaskFig}\footnotesize%
Wiener-filtered QMASK map combining the QMAP and Saskatoon experiments. The CMB
temperature is shown in coordinates where the north celestial pole is at the
center of the dashed circle of $16\arcdeg$ diameter, with R.A. being zero at the
top and increasing clockwise.
This map differs from the one published in \protect\cite{qmask}
by the erasing of QMAP information for $\l\simgt 200$ described
in the text.
}
\end{figure}

Since the accuracy of recent 
measurements leftward of the first acoustic peak is
limited by sample-variance rather than instrumental noise, 
we will use the largest area CMB map publicly available to date with degree scale
angular resolution. This map, nicknamed QMASK \cite{qmask}, is shown 
in \fig{newqmaskFig} and combines
the data from the QMAP \cite{qmap1,qmap2,qmap3} and 
Saskatoon \cite{Netterfield95,Netterfield97,saskmap} (hereafter SASK sometimes) experiments 
into a 648 square degree map around the North Celestial Pole.
This map has been extensively tested for systematic errors \cite{qmask},
with the conclusion that the QMAP and Saskatoon
experiments agree well overall, as well as analyzed for non-Gaussianity
\cite{genus,minkowski}.
However, for the present power spectrum analysis,
it is important to perform additional systematic tests to see if there is evidence
of scale-dependent problems in any of the maps.
In particular, pointing problems, beam uncertainties, sampling and pixelization effects
can smear the maps in a way that changes the shape of the power spectrum, 
suppressing small-scale fluctuations (and under some circumstances boosting the
fluctuations as well).

The rest of this paper is organized as follows.
In Section~\ref{CombiningSec},
we present a technique for erasing this type of unreliable information,
and apply it to produce a new combined QMASK map where all the statistical
weight on small scales comes from Saskatoon.
We compute the power spectrum of this combined map 
in Section~\ref{SpectrumSec}. We discuss systematic errors in
section~\ref{DiscSec},
finding good agreement but a hint of suppressed 
QMAP power for $\l\simgt 200$, and conclude with a rather conservatively
cut data set that we believe to be reliable as a starting point for 
cosmological parameter analysis.


\section{Combining the SASK and QMAP Experiments}
\label{CombiningSec}

As will be discussed in more detail in \sec{DiscSec}, pointing problems, 
beam uncertainties, pixelization and sampling effects
can suppress small-scale fluctuations in the CMB power spectrum. The QMAP
results may therefore only be valid on large angular scales 
\cite{qmap1,qmap2,qmap3}. 
Both pointing inaccuracies and pixelization effects could have effectively smoothed
the QMAP map, suppressing small-scale power.
Flight 1 of QMAP \cite{qmap1} was sampled at a relatively low rate, causing the effective
beam shapes to be elongated along the scan direction. Since this ellipticity
was not modeled in the mapmaking algorithm \cite{qmap3}, the resulting 
smoothing would again be expected to suppress small-scale power.
Miller {\etal} \cite{Miller01} present a detailed discussion of these issues,
and conclude that the QMAP measurements are likely to be unreliable for
$\l\simgt 200$.
Although attempts can be made to model and correct for some of these effects,
an accurate treatment of the undersampling problem in particular would be way
beyond the scope of the present paper, requiring the entire maps to be regenerated
from the time-ordered data with an order of magnitude more pixels to be able to
resolve the beam ellipticity (the ellipticity is not uniform in direction, since 
the sky rotated relative to the scan axis during the flights). Such an analysis
would hardly be worthwhile anyway, since the strength of QMAP compared to subsequent 
experiments lies on large scales, not on small scales.

For these reasons, we adopt a more conservative approach, 
combining the QMAP and Saskatoon maps in 
such a way that the small-scale QMAP information can be optionally erased
as a precaution.
By this we do not mean removing the {\it signal} 
(smoothing the map), which would just lead to further underestimation of the
true power. Rather, we mean removing the {\it information}, \ie, doing
something that causes subsequent analysis steps (like combining with Saskatoon
or measuring the power spectrum) to give negligible statistical weight to the
small-scale QMAP signal. We achieve this by creating a random map with very
large small-scale noise and adding it to the QMAP map, modifying its noise
covariance matrix $\N$ accordingly.

In practice, we start by generating a white noise map $\xwhite$ which has the following
properties: it covers the same sky region as QMAP, and each pixel temperature is 
drawn independently from a Gaussian distribution with zero mean and 
standard deviation $\sigma$, giving it a noise covariance
matrix $\Sig_{white}=\sigma^2\I$.
This makes its angular power spectrum $C_\l$ independent of $\l$.
We then apply the Laplace operator $\nabla^2$ to the mock map.
Since it is pixelized on a square grid, we  do this in practice
by multiplying by a matrix $\L$ that subtracts each pixel from the average of
its four nearest neighbors. 
The transformed map $\xblue\equiv\L\xwhite$ thereby obtains a very blue power
spectrum $C_\l\simpropto {\ell}^4$,
since Laplace transformation corresponds to multiplying by
$\l(\l+1)$ in the Fourier (multipole) domain.
We choose the normalization factor $\sigma$ such that 
the blue noise starts dominating the noise and signal of the QMAP map
around $\l\sim 200$. Since the CMB power falls off as 
$C_\l\simpropto {\ell}^{-2}$, the result is that 
the added noise is negligible for $\l < 200$ and dominates 
completely for $\l > 300$.
Finally, we add this blue noise map to the QMAP map, obtaining
\beqa{fixedQMAPEq}
\x_{new}&=&\x_{\rm QMAP} + \xblue = \x_{\rm QMAP} +\L\xwhite,\\
\Sig_{new}&=&\Sig_{\rm QMAP}+\Sblue = \Sig_{\rm QMAP}+ \sigma^2\L\L^t.
\eeqa
To quantitatively assess the effect of this procedure, we perform a series of numerical 
experiments where we rescale the noise level by a factor $p=0,1,3,5,7,1000$.
This corresponds to multiplying
$\xblue$ by $p$ and multiplying $\Sblue$ by $p^2$.
${\it p}=0$ means adding no noise, \ie, that $\x_{new}$ retains
all QMAP information; $p=\infty$ (or $p=1000$ in practice) means
that $\x_{new}$ contains essentially pure noise. 
Below we will see that the choice $p=3$ erases information on
small scales $\l>200$ very well, while doing little harm on 
larger scales. After erasing with $p=3$, we 
combine the resulting maps with the Saskatoon data as in \cite{qmask}.
The result is shown in \fig{newqmaskFig}, and looks almost unchanged 
(compare Figure 4 in \cite{qmask})
since 
Wiener filtering suppresses noisy modes and the smallest scales were noisy
to start with.

\section{The Angular Power Spectrum}
\label{SpectrumSec}

\begin{figure}[tb]
\vskip-1.4cm
\noindent\centerline{\epsfxsize=9.5cm\epsffile{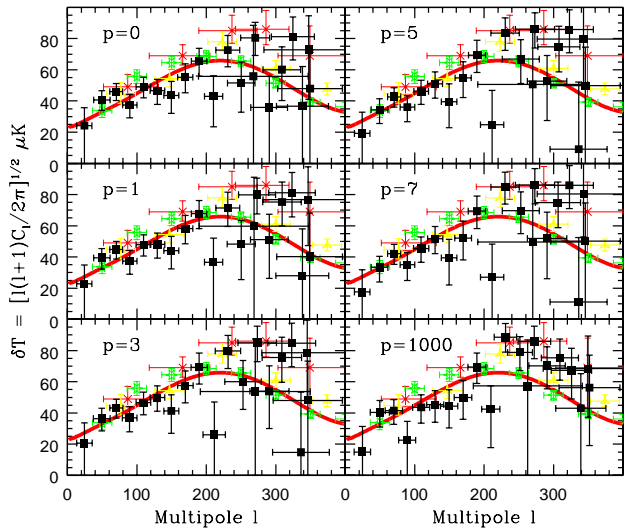}}
\vskip-1cm
\smallskip
\caption{\label{Spectrum2Fig}\footnotesize%
Angular power spectrum $\delta T\equiv[\l(\l+1)C_\l /2\pi]^{1/2}$ 
(uncorrelated) in 20 bands of CMB anisotropy from the combined QMASK data in case of ${\it
p}=0$(no additional noise at all, just simple combination of SASK and QMAP),1,
3, 5, 7 and 1000 (no QMAP information, only SASK information). For
comparison, we also plot the a recent ``concordance'' model 
\protect\cite{concordance}
and the power measurements from 
MAXIMA and BOOMERanG.
}
\end{figure}

In this section, we compute the angular power spectrum of the QMASK map produced
in the previous section, shown in \fig{newqmaskFig}.
It contains 6495 pixels and covers a 648 square degree sky region. 
We calculate the angular power spectrum  
using the quadratic estimator method of 
\cite{cl,BJK}, implemented as described in 
 \cite{Padmanabhan99,polarization}.
This method involves the following steps: 
(i) S/N compression of data and relevant matrices by omitting 
Karhunen-Lo\`{e}ve (KL) eigenmodes with
very low signal-to-noise ratio, 
(ii) computation of Fisher matrix and raw quadratic estimators,
(iii) decorrelation of data points. 
We compute the power in 20 bands from $\l=2$ to 400 of width $\Delta\l=20$,
which takes about a week on a workstation.
We repeated this calculation in its entirity for 
the following values of the $p$-parameter:
0 (no information erased), 1, 3, 5, 7 and 1000 (Saskatoon only, QMAP fully erased),
as shown in \Fig{Spectrum2Fig}.

\begin{figure}[tb]
\vskip-1.4cm
\noindent\centerline{\epsfxsize=9.5cm\epsffile{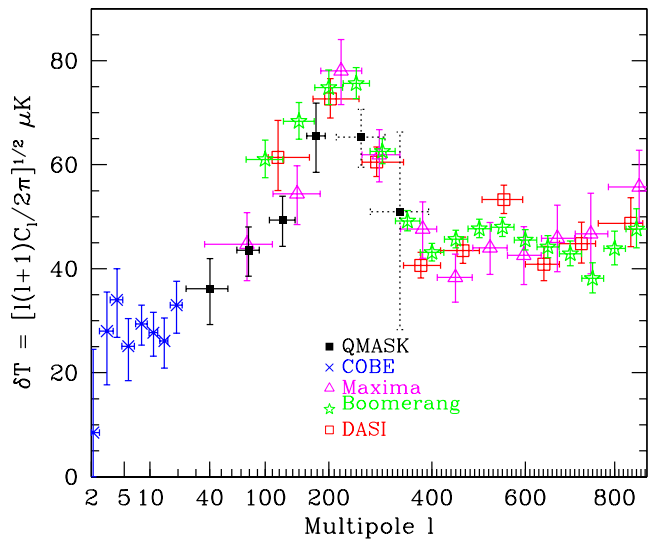}}
\vskip-1cm
\smallskip
\caption{\label{SpectrumFig}\footnotesize%
Uncorrelated measurements of the CMB power spectrum 
$\delta T\equiv[\l(\l+1)C_\l /2\pi]^{1/2}$ 
from the combined QMASK data. 
For comparison, we also plot 
the measurements from 
COBE/DMR, MAXIMA, DASI and BOOMERanG. 
These error bars do not include calibration uncertainties 
of 10\% (QMASK), 10\% (BOOMERanG), 4\% (DASI) and 4\% (MAXIMA).
As described in the text, we suggest using only the first four points 
$(\l\simlt 200)$, not the two dashed ones, for cosmological model constraints.
}
\end{figure}

In the progression of power spectra shown in this figure,
the $\l$-values beyond which QMAP is effectively erased is seen to 
shift to the left as $p$ increases. For $p=3$, this scale is seen to be
of order 200 in the sense that the leftmost points are almost 
indistinguishable from the $p=0$ case whereas those with decent signal-to-noise
for $\l\simgt 200$ (say points number 10 and 11) are similar to the $p=1000$ case.
Adopting this cutoff scale as our baseline calculation (we further discuss this
choice in the next section),
we then average these rather noisy $p=3$ measurements 
into six (still uncorrelated)
measurements listed in Table 2. The first four probe angular scales where
the above-mentioned systematics are likely to be negligible, 
incorporating the first 9 $\l$-bands (up to $\l=180$),
and are shown in \fig{SpectrumFig}.
The horizontal bars show the mean and rms width of the corresponding
window functions.

\begin{center}
\begin{tabular}{|c|c||c|c|} 
\hline
\multicolumn{2}{|c||} {${\it p}=0$} & \multicolumn{2}{c|} {${\it p}=3$}\\
\hline
$\l$  & $\delta T_\l^2 \> [\mu K^2]$ & $\l$ & $\delta T_\l^2 \> [\mu K^2]$\\ 
\hline
$   24\pm 11$  & $   641\pm    763 $ &$  24 \pm 11 $ &  $   474\pm   769$\\
$   49\pm 12$  & $  1821\pm    559 $ &$  49 \pm 12 $ &  $  1459\pm   570$\\
$   70\pm  9$  & $  2327\pm    551 $ &$  70 \pm  9 $ &  $  2056\pm   567$\\
$   90\pm 10$  & $  1543\pm    606 $ &$  90 \pm 10 $ &  $  1492\pm   629$\\
$  110\pm 10$  & $  2646\pm    698 $ &$ 109 \pm 10 $ &  $  2403\pm   738$\\
$  130\pm 11$  & $  2386\pm    815 $ &$ 130 \pm 11 $ &  $  2737\pm   885$\\
$  150\pm 11$  & $  2083\pm    955 $ &$ 150 \pm 11 $ &  $  1872\pm  1071$\\
$  170\pm 11$  & $  3358\pm   1108 $ &$ 170 \pm 11 $ &  $  3652\pm  1278$\\
$  190\pm 11$  & $  4717\pm   1266 $ &$ 190 \pm 11 $ &  $  5265\pm  1487$\\
$  210\pm 13$  & $  2030\pm   1430 $ &$ 211 \pm 13 $ &  $   749\pm  1689$\\
$  231\pm 17$  & $  5791\pm   1586 $ &$ 231 \pm 17 $ &  $  7007\pm  1861$\\
$  250\pm 20$  & $  2918\pm   1728 $ &$ 253 \pm 20 $ &  $  3964\pm  1991$\\
$  270\pm 21$  & $  7116\pm   1865 $ &$ 273 \pm 21 $ &  $  7952\pm  2101$\\
$  291\pm 27$  & $  1408\pm   2009 $ &$ 291 \pm 27 $ &  $  3207\pm  2211$\\
$  309\pm 27$  & $  3969\pm   2182 $ &$ 309 \pm 27 $ &  $  6268\pm  2352$\\
$  325\pm 31$  & $  7263\pm   2421 $ &$ 323 \pm 31 $ &  $  7903\pm  2563$\\
$  339\pm 37$  & $  1475\pm   2776 $ &$ 336 \pm 37 $ &  $ 241.4\pm  2897$\\
$  349\pm 47$  & $  2530\pm   3278 $ &$ 346 \pm 47 $ &  $  2540\pm  3378$\\
$  348\pm 65$  & $  5842\pm   3985 $ &$ 345 \pm 65 $ &  $  6800\pm  4064$\\
$  268\pm 79$  & $  3430\pm   5187 $ &$ 271 \pm 79 $ &  $  3167\pm  5237$\\
\hline 
\end{tabular}
\end{center}

{
\footnotesize 
{\bf Table 1.} The power spectra $\delta T\equiv[\l(\l+1)C_\l
/2\pi]^{1/2}$ from the QMASK map in two cases:${\it p}=0$ and ${\it p}=3$. The
tabulated error bars are uncorrelated between  the twenty measurements, but do
not include an overall calibration uncertainty of 10\% for $\delta T$.
 }

\begin{center}
\begin{tabular}{|c|c|} 
\hline 
$\l$  & $\delta T_\l^2 \> [\mu K^2]$ \\ 
\hline 
$ 40\pm 16$& $1308\pm 452$ \\
$ 79\pm 13$& $1899\pm 410$ \\
$125\pm 19$& $2436\pm 471$ \\
$178\pm 15$& $4295\pm 869$ \\
\hline
$259\pm 46$& $4265\pm 726$ \\
$335\pm 58$& $2596\pm 1795$ \\
\hline 
\end{tabular}
\end{center}
{\footnotesize
{\bf Table 2.} The power spectrum $\delta T\equiv[\l(\l+1)C_\l /2\pi]^{1/2}$
from the QMASK map. The tabulated error bars are uncorrelated between 
the six measurements, but do not include an overall calibration uncertainty
of 10\% for $\delta T$. We recommend using only the first four for
cosmological model constraints.
}
\label{DeltaTab}


\section{Discussion}
\label{DiscSec}

We have measured the CMB power spectrum from the combined QMAP and Saskatoon
data sets, obtaining significant detections in the range $\l\sim 30-300$.
The key question that we need to address in this section is how reliable these 
measurements are.

The ``usual suspects'' as far as systematics and non-CMB contamination are concerned 
tend to have power spectra that are either redder or bluer than that of the CMB,
and therefore naturally split into two categories:
\begin{itemize}

\item Problems on {\it large angular scales} can be caused by
contamination from diffuse foregrounds (synchrotron, free-free and spinning dust emission),
and systematic errors related to atmospheric contamination, scan-synchronous offsets,
atmospheric contamination, {\etc}

\item Problems on {\it small angular scales} can be caused by
contamination from point source foregrounds
and systematic errors related to 
pointing problems, beam uncertainties,  pixelization and sampling effects.

\end{itemize}

The foreground contamination \cite{foregpars} has been previously quantified for both 
Saskatoon \cite{saskforeg} and QMAP \cite{qmapforeg}, by cross-correlating the
maps in question with a foreground templates tracing synchrotron, dust and
free-free emission,
point sources, {\etc}
The conclusion was that foregrounds contribute at most a few percent to the angular
power spectrum reported here, mainly on the largest angular scales.
The errors reported on the power spectrum
assume that the underlying CMB signal is Gaussian, which is supported
by two recent Gaussianity analyses of the QMASK map \cite{genus,minkowski}.

As opposed to foregrounds, which are true features on the microwave sky, 
the remaining problems listed above are experiment-specific effects which 
would be expected to affect QMAP and Saskatoon differently.
This provides us with a powerful tool with which to test for their presence, which
we will now employ:
comparing the QMASK and Saskatoon maps where they overlap. 

Previous comparisons between CMB data sets have been done either spatially
or in terms of power spectra (discarding phase information). 
Since two inconsistent maps can have identical power spectra, 
one should be able to obtain still stronger tests for systematic errors
by comparing power {\it with} phase information. 
For instance, one could imagine band-pass filtering the two maps 
to retain only a particular range of multipoles $\l$, and then testing 
whether these two filtered maps were consistent. We will now describe
a simpler way of implementing a test in this spirit.

\subsection{A method for scale-by-scale comparison of two maps}

\label{testSec}

Numerous map comparisons have been performed with the
``null-buster'' test \cite{comparing}
\beq{devEq}
  \nu \equiv \frac{\z^t\N^{-1}\SS\N^{-1}\z - \tr\{\N^{-1}\SS\}}
  {\left[2\>\tr\{\N^{-1}\SS\N^{-1}\SS\}\right]^{1/2}},
\eeq
where 
$\nu$ can be interpreted as the number of ``sigmas'' at which
the difference map $\z$ is inconsistent with pure noise.
If the two maps are stored in vectors $\x_1$ and $\x_2$ and have noise
covariance matrices $\N_1$ and $\N_2$, then a weighted difference map
$\z\equiv\x_1-\r\x_2$ will have noise covariance 
$\N\equiv\N_1 + r^2\N_2$. The matrix $\SS$ tells the test which 
modes (linear combinations of the pixels $\z$) to pay most attention to, 
and can be chosen arbitrarily.
The choice $\SS=\N$ gives a standard $\chi^2$-test. It can be shown \cite{comparing}
that the null hypothesis that $\z$ is pure noise (that $\expec{\z\z^t}=\N$)
is ruled out with maximal significance on average if $\SS$ is chosen to be the covariance
of the expected signal in the map, \ie, $\SS=\expec{\z\z^t}-\N$.
In our case, we choose $\SS$ to be the CMB covariance matrix corresponding to 
a power spectrum
\beq{deltaEq} 
\delta T_\l = \left\{\begin{array}{ll} \mbox{1 $\muK$ } &
\mbox{ if $\l \in [\l _{min},\l _{max}]$};\\ \mbox{0 $\muK$} & \mbox{
otherwise}. \end{array}\right., 
\eeq
ensuring that the test only uses information in the multipole 
interval $[\l _{min},\l _{max}]$. 
The overall normalization of $\SS$ is irrelevant, since it 
cancels out in \eq{devEq}.

\subsection{Results of comparing QMAP and Saskatoon scale-by-scale}
\label{TestSub}

We will now use the method described above to compare the QMAP and Saskatoon data scale-by-scale, 
\ie, in different multipole intervals. 

The QMASK map has been shown to be inconsistent with noise at the $62\sigma$ 
level \cite{qmask}. In the region where the QMAP and Saskatoon maps overlap, they were 
found to detect signal at $40\sigma$ and $21\sigma$, respectively, while the
difference map was consistent with pure noise.
Which angular scales are contributing most of this information,
and how well do the two maps agree scale-by-scale?

\begin{figure}[tb] \vskip-1.7cm 
\vskip-0.2cm 
\centerline{\epsfxsize=8.5cm\epsffile{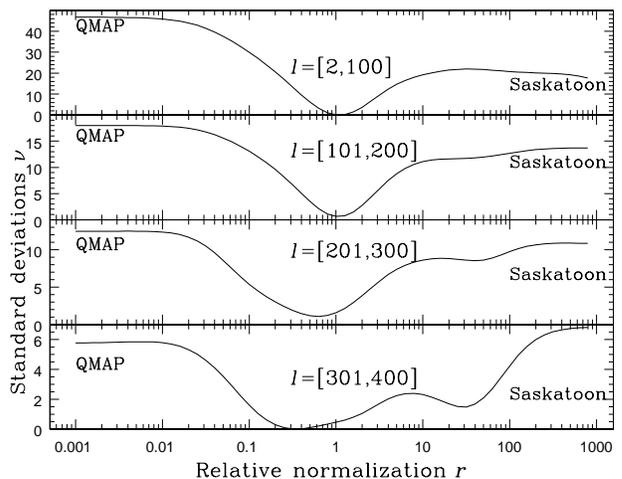}} 
\vskip0.2cm 
\caption{\label{ClslopeFig}\footnotesize%
Comparison of the QMAP and
Saskatoon experiments on different angular scales, corresponding to
the multipole ranges shown in square brackets.
The curves show
the number of standard deviations (``sigmas'') at which the difference map
$\xt_{ \rm QMAP}-r\xt_{\rm SASK}$ is inconsistent with mere noise. Note that
this is only for the spatial region observed by both experiments.
} 
\end{figure}

To answer these questions,  \Fig{ClslopeFig} shows the result of comparing
QMAP with Saskatoon in the four multipole intervals  $[2,100]$, $[101,200]$,
$[201,300]$, and $[301,400]$. QMAP is seen to detect signal in the overlap
region at  $47\sigma$, $18\sigma$, $12\sigma$ and $6\sigma$, respectively,
whereas the corresponding numbers for Saskatoon are  $18\sigma$, $14\sigma$,
$11\sigma$ and $7\sigma$. In other words, QMAP dominates on large scales,
whereas Saskatoon has at least comparable information 
content on small scales because of superior angular resolution.

Although QMAP and Saskatoon both detect significant CMB signal in all four
bands, this signal is seen to be common to both maps since the difference maps
$\z$ for $r=1$ are consistent with noise. Furthermore, there is no evidence of
relative calibration errors for the $[2,100]$ or $[101,200]$ bands, since the
minima of these two curves are at $r\approx 1$. However, the situation is
less clear on smaller scales: the best-fit amplitude of QMAP is only $63\%$
of the amplitude of Saskatoon for $\ell\in [201,300]$, and even lower for
$\ell\in [301,400]$. None of these departures of the minimum from $r=1$ are
statistically significant --- we cannot determine whether this is a a problem 
or not simply because the amount of information in the maps drops sharply on
small scales where detector noise and beam dilution become important.
However, whereas the Saskatoon information was extracted from highly
oversampled calculations of the relevant beam patterns on the sky, 
and should be reliable on small scales,
there are a number of reasons why the QMAP
data may only be valid on larger ($\l\simlt 200$) angular scales 
\cite{qmap1,qmap2,qmap3}:
\begin{enumerate}
\item The QMAP pointing solution was only accurate to this level \cite{qmap2},
and small residual pointing uncertainties could have effectively smoothed
the map, suppressing power for $\l\simgt 200$
relative to Saskatoon just as \fig{ClslopeFig}
indicates.
\item The QMAP
maps were generated by subdividing the sky into square pixels of side
$\theta=0.3125^\circ$, and the effect of this pixelization may well become
important on angular scales substantially exceeding $\l\sim 1/\theta\approx
200$, suppressing power on these scales.
\item Flight 1 of QMAP \cite{qmap1}, which dominates the sky coverage in 
\fig{newqmaskFig} was sampled at a relatively low rate, causing the effective
beam shapes to be elongated along the scan direction. Since this ellipticity
was not modeled in the mapmaking algorithm \cite{qmap3}, the resulting 
smoothing would again be expected to suppress power on scales $\l\simgt 200$.
\end{enumerate}
Miller {\etal} \cite{Miller01} present a detailed discussion of these issues,
and conclude that the QMAP measurements are likely to be unreliable for
$\l\simgt 200$. This justifies our $p=3$ choice in Section 2, effectively discarding
this suspect small-scale information from QMAP.

\subsection{Conclusions}

We have measured the CMB power spectrum from the combined QMAP and Saskatoon
data sets, and performed a number of tests for potential systematic errors.
On large angular-scales, $\l\simlt 200$, our results appear very solid:
foreground contamination has been quantified to contribute no more than a few percent 
to the power spectrum, and we find 
beautiful internal consistency between the QMAP and Saskatoon components of our map,
both in terms of power amplitude and in terms of spatial phase information.
On smaller scales $\l\simgt 200$, there are good {\it a priori} reasons to 
discard the QMAP information, and we have therefore done so.
Our internal scale-by-scale comparison between QMAP and Saskatoon independently 
suggests that there may be systematic problems on these scales.
Since this unfortunately leaves us with no independent way of validating 
our Saskatoon results in this small-scale regime, \eg, testing our Saskatoon deconvolution
procedure, we strongly recommend the conservative approach of using only the
first four band power measurements we have reported in table 2 and Figure 3 --- this
is also where QMASK is most sensitive relative to other experiments.
These four measurements, spanning the range $\l\sim 30-200$, 
provide among the sharpest and best tested constraints to date on 
these scales, provide a good starting point for constraining cosmological models.

\subsection{Comparison with other experiments}

\Fig{Spectrum2Fig} shows that 
our results agree well with the ``concordance'' model of \cite{concordance}.
They are also consistent with BOOMERanG \cite{Netterfield01}, DASI \cite{Halverson01}
and Maxima \cite{Hanany00} once calibration uncertainties are
taken into account. The QMAP and Saskatoon calibration uncertainties
are $6\%-10\%$ and $10\%$, respectively (we have corrected the original $\delta T$ results 
\cite{qmap1,Netterfield97} by a factor 1.05 using the latest Cassiopeia A 
data \cite{Mason99} as in \cite{Gawiser00}). Although
the uncorrelated components of this may average down somewhat when the two maps
are combined, we quote a $10\%$ uncertainty on our result to be conservative.
Our results are also consistent with those obtained from QMAP and Saskatoon
alone. Our leftmost data point agrees well with the last point
measured from COBE/DMR \cite{Bennett96} by \cite{cl}.

The original BOOMERanG98 power spectrum was based on a map area of 
436 square degrees \cite{deBernardis00}, so one might expect our error bars
on large scales to be a factor $(648/436)^{1/2}\approx 1.2$ smaller.
Our actual error bars on $\delta T_\l^2$ are only about 10\% smaller 
on the best QMASK scales after adjusting for bandwidth
differences, which is because of 
BOOMERanG's lower noise
levels (the scan strategy and $1/f$-noise of QMAP 
introduced a non-negligible amount on
noise even on the largest scales). 
We note that the substantial reduction in error bars relative to 
the original Saskatoon analysis \cite{Netterfield97} is due to additional
information not only from QMAP, but from Saskatoon as well.
This is because our present method extracts all the information present,
whereas that employed in \cite{Netterfield97} was limited to information along 
radial scans, not using phase information between scans.


In conclusion, we have measured the CMB power spectrum
on angular scales $\l\sim 30-200$ from the QMASK map, 
placing among sharpest and best tested constraints to date
on the shape of the CMB power spectrum as it rises towards the 
first acoustic peak. Our window functions, the combined
map and its noise covariance matrix are available at\\ 
$www.hep.upenn.edu/\sim xuyz/qmask.html$.


\bigskip
The authors wish to thank Mark Devlin, Lyman Page 
and an anonymous referee for useful comments. 
Support for this work was provided by
NSF grant AST00-71213,
NASA grants NAG5-9194 and NAG5-11099,
the University of Pennsylvania Research Foundation,
the Zaccheus Daniel Foundation and the David and Lucile Packard Foundation.




\end{document}